# The signatures of conscious access and phenomenology are consistent with large-scale brain communication at criticality

Enzo Tagliazucchi

Netherlands Institute for Neuroscience, Meibergdreef 47, 1105 BA Amsterdam-Zuidoost, The Netherlands

Corresponding author: tagliazucchi.enzo@googlemail.com

**Abstract**

Conscious awareness refers to the association of information processing in the brain that is accompanied by subjective, reportable experiences. Current models of conscious access propose that sufficiently strong sensory stimuli ignite a global network of regions allowing further processing. The immense number of possible experiences indicates that brain activity associated with conscious awareness must be highly differentiated. However, information must also be integrated to account for the unitary nature of consciousness. We present a conceptual computational model that identifies conscious access with self-sustained percolation in an anatomical network. We show that if activity propagates at the critical threshold, the amount of integrated information (Φ) is maximal after conscious access, as well as other related markers. We also identify a posterior "hotspot" of regions with high levels of information sharing during conscious access. Finally, competitive activity spreading qualitatively describes the results of paradigms such as backward masking and binocular rivalry.

## 1. Introduction

Consciousness is both a mystery and a commonplace experience. The most daunting question about consciousness is how a particular portion of matter - the brain - can develop a unique, first-person point of view, phenomenal experiences, and even reflect upon its own nature (Chalmers, 1995). An associated question is why phenomenal experiences feel the way they feel; in other words, what properties of such portion of matter account for the *redness* of the red, or the suffering associated with a toothache (Lewis, 1956). Philosophers have battled with these difficult questions (and also among themselves) for centuries, leading to the contemporary consensus that there is indeed something to be explained scientifically about consciousness, and that such explanation is most likely to involve the brain (Dennett, 1993).

The contemporary neuroscience of consciousness builds upon this consensus and suggests the experimental approach of finding the neural correlates of a conscious experience, i.e. the minimal set of events in the brain (presumably the firing of certain neurons) necessary for such a conscious experience to occur. This is the backbone of the research programme proposed by Bachmann (Bachmann, 1984) and afterwards by Crick & Koch in their influential 1990 article (Crick & Koch, 1990). A wealth of experimental data on the neural correlates of different sensory modalities has accumulated since Crick & Koch's seminal proposal, which will not be presented here for reasons of space, but can be read from a number of comprehensive review articles (Dehaene, 2014; Bachmann, 2015; Koch et al., 2016a). In recent years, a shift has occurred in the research community from the search of neural correlates of consciousness towards the search of models of consciousness (Seth, 2007). A model of consciousness is understood here as a mechanistic (either qualitative or quantitative) explanation of how brain activity leads to conscious awareness. Ideally, models of consciousness must have predictive power, i.e. they should be able to predict from their proposed mechanism what the neural correlates of a given conscious experience are.

An influential model of conscious access[1] has been proposed by Baars, Changeux and Dehaene and is commonly referred to as the global workspace model (Baars, 2005; Dehaene et al., 1998; Dehaene et al., 2011; Dehaene & Changeux, 2011). Their proponents suggest a competitive mechanism at a neural periphery consisting of parallel processors. A non-linear, all-or-none transition ("ignition") occurs when one of the competing incoming stimuli gains access to a distributed set of neurons comprising the global workspace. Such transition leads to self-sustained activity that can be accessed by different brain processes (e.g. working memory, decision making).

---

[1] In this work, "conscious access" refers to the functionalist or psychological perspective on consciousness. "Conscious phenomenology" refers to the subjective, first-person qualities of consciousness (see the discussion in Chalmers, 1996, for further clarification on this distinction). "Conscious awareness" is more loosely employed to refer to conscious access and the associated phenomenology going hand by hand (but see Block, 2005). Finally, "conscious state" refers to a state in which conscious awareness can occur (e.g. wakefulness as opposed to the comatose state), which is not necessarily disentangled from conscious content itself (Bachmann & Hudetz, 2014).

The competition to gain access to the global workspace accounts for the results of several psychophysical experiments, such as backward masking and binocular rivalry – i.e. for a short time after its ignition by a given stimuli, the global workspace remains inaccessible to other stimuli (Dehaene et al., 2006). Neuroimaging experiments have consistently identified the global workspace with the fronto-parietal cortex of the brain (for an example, see Carmel et al., 2006; further examples are discussed in the references on the neural correlates of consciousness). Thus, it can be said that the global workspace model is one of conscious access: it posits that the function of conscious awareness is the broadcasting of information in the brain, and suggests a plausible neuronal mechanism for such broadcasting.

Other models of consciousness address instead the phenomenology of consciousness. The information integration theory put forward by Giulio Tononi identifies the properties that a physical system should have in order to present a high level of differentiation, as well as of information integration. Differentiation relates to the phenomenal experience of the uniqueness of any conscious event, each a single one among an astronomical number of possible conscious experiences. While differentiation is maximized by statistically independent neuronal firing, the unitary nature of each conscious experience suggests that information integration is another key physical mechanism underlying conscious phenomenology. From these considerations, Tononi and colleagues have derived a series of mathematical algorithms to quantify the level of information integration in an arbitrary physical system ($\Phi$) (Tononi, 2004; Tononi, 2011; Tononi et al., 2016). While generally intractable, heuristics and approximations have been employed to support the theory (Barrett & Seth, 2011; Sasai et al., 2016). It is nevertheless important to remark that a high value of $\Phi$ is proposed as indicative of consciousness in a given physical system, but does not necessarily specify the mechanisms that generate such large value of $\Phi$.

This description of both models suggests that the global workspace theory and the information integration theory address two different aspects of the conscious experience: the first tackles the issue of conscious access (closely related to the function of conscious awareness), while the second the phenomenology of consciousness. The objective of the present work is to propose a mechanism based on the global workspace model that is compatible with the proposal that high values of $\Phi$ are indicative of conscious awareness. In doing so, we also put forward a mechanism based on simple physical principles by virtue of which a system can attain high values of $\Phi$. In other words; we aim to show that both theories could be seen as compatible when considered as theories of conscious access and the phenomenology of consciousness, respectively.

The mechanism we propose to fulfill our objective is the propagation of information in the brain at the "edge of failure", i.e. at the critical point at which activity becomes self-sustained (Haimovici et al., 2013). This mechanism is similar to that proposed by a class of mathematical models of the cerebral cortex known as neuropercolation models (Kozma, 2007). As we will discuss extensively below, there is empirical evidence supporting critical-like dynamics in the human brain, adding experimental plausibility to the proposed mechanism (Chialvo, 2010). We develop a conceptual computational model incorporating realistic anatomical connectivity to test the feasibility of our hypothesis, and then discuss the relationship between our results and more neurobiologically realistic models.

2. **Materials and Methods**

*2.1 Anatomical connectivity network*

As part of our computational model we employ an anatomical connectivity network of the cerebral cortex inferred from diffusion tensor imaging (DSI) data from 5 healthy participants (mean age 29.4 years, all male) (the "anatomical connectome") (Hagmann et al., 2008). The network comprises 998 regions of interest (each 1.5 cm$^2$ in area) placed throughout the cortex (but excluding sub-cortical structures and the cerebellum) of individual participants, being later mapped into a common space. White matter tractography was applied to compute fiber trajectories and to construct a network by linking every two nodes for which a fiber existed starting in one and ending in the other. A network representation was obtained by averaging the fiber densities between nodes of all participants. The weighted adjacency matrix of this average network is noted as $W_{ij}$.

*2.2 Computational model*

The model is a variant of the Greenberg-Hastings cellular automaton of excitable dynamics (Greenberg & Hastings, 1978). Each node in the network can be, at any given time, in one of three possible states: inactive, active, and refractory. The rules governing the transitions between these states are as follows:

1) If inactive, the i-th node becomes active if $\sum_{j\ is\ active} W_{ij} > T$. The threshold $T$ determines how difficult is for the activity to spread in the network (i.e. very high $T$ makes spread unlikely). Note that this model does not include the possibility of spontaneous activation.

2) If active, a node always transitions towards the refractory state in the next time step.

3) A node in the refractory state transitions back towards the inactive state with a probability of $10^{-1}$.

All three rules are applied synchronously. This model has only two parameters: the threshold $T$ and the probability of transitioning from the refractory to the inactive state. Such probability represents the level of inhibition in the system, since a refractory node is immune to activation. We held this probability constant and varied the threshold $T$. It has been shown that the model presents a non-linear phase transition for a given value of $T$, which we note as $T_C$ (Copelli & Campos, 2007; Fates, 2010; Haimovici et al., 2013). Below $T_C$ the activity cannot become self-sustained and is extinguished, and at $T_C$ activity becomes self-sustained at the edge of extinction. The biological interpretation of the parameter $T$ is that of the stimulus strength in combination with the dynamical and structural properties behind its propagation in the cerebral cortex.

*2.3 Initial conditions and competitive dynamics*

To simulate the spread of activity towards the global workspace from an incoming sensory stream, all simulations started with only one node at the active state (except when simulating parallel competition; see below).

Competition between two stimuli was modeled by simultaneously simulating two cellular automata following the rules described in the previous section, with the additional constraint that an active or refractory node associated with one stimuli "blocks" its activation by the second stimuli (in the case that the activation conditions are met). In other words, an already active or refractory node cannot be activated by the propagation of the other stimuli. In the most general case, each competing automata had its own propagation threshold, $T_1$ and $T_2$. Competing dynamics were investigated serially (i.e. the same node was activated twice after a certain number of time steps) and in parallel (i.e. two nodes were activated simultaneously, each activation propagating with its own threshold value).

*2.4 Mutual information*

We measured mutual information $I(X;Y)$ between two simulated signals $X, Y$ as follows:

$$I(X;Y) = \sum_{x,y} P_{XY}(x,y) \, log \, \frac{P_{XY}(x,y)}{P_X(x)P_Y(y)}$$

**[1]**

In this equation, $X$ and $Y$ are treated as two random variables which can assume three different values identified with the lower-case letters $x$ and $y$; these values correspond to the active, inactive and refractory states. $P_{XY}(x, y)$ is the joint probability distribution of $X$ and $Y$; and $P_X(x)$ and $P_Y(y)$ are the marginal probability functions of $X$ and $Y$, respectively.

Motivated by a previous study showing that the level of long-range information sharing indexes levels of consciousness in patients (King et al., 2013), we obtained the average mutual information between all pairs of nodes separated by less than the mean Euclidean distance between all nodes (i.e. short-range average mutual information) and by more than this mean (i.e. long-range average mutual information).

*2.5 Algorithmic complexity*

The level of algorithmic complexity of brain activity has been proposed by Tononi and colleagues as an indirect marker of the level of consciousness (Casali et al., 2013). The algorithmic complexity (or Kolmogorov-Chaitin complexity) of a binary string is defined as the minimum length of a computer program that reproduces such string (theoretically, the computer program is assumed to be implemented in a universal Turing machine) (Chaitin, 1977). The algorithmic complexity of a string is incomputable, i.e. it is impossible to devise an algorithm that outputs the algorithmic complexity of an arbitrary binary string. However, algorithms approximating the algorithmic complexity of strings have been developed in the context of information compression. The Lempel-Ziv algorithm outputs the level of (lossless) compressibility of a string (Ziv & Lempel, 1977), and has been employed on brain activity recordings as a marker of conscious awareness (Casali et al., 2013; Abasolo, 2015).

To apply the Lempel-Ziv algorithm to the signals generated by our model, we first binarized them using the convention of 1 = active and 0 = inactive or refractory. Since the algorithmic complexity is biased by the number of 0's and 1's in the time series, we shuffled each signal 100 times and computed the algorithmic complexity of each shuffled signal. We then divided the value obtained for the unshuffled signal by the average complexity of all shuffled signals.

*2.6 Information integration (Φ)*

The latest versions of Φ are of difficult practical application (Tononi, 2011). We therefore considered a more tractable version of Φ introduced by Barrett and colleagues (Barrett & Seth, 2011), and recently applied to experimental fMRI data and to the results of a computational model (Sasai et al., 2016, Mediano et al., 2016).

This definition of Φ is based on how well a system $X$ can predict its own future (i.e. decoding its own past) after τ time steps, when considered as a whole vs. as a sum of two subsystems, $X^1$ and $X^2$.

This computation evaluates how much information the whole system $X$ generates in excess of the individual contributions of $X^1$ and $X^2$. Considering an arbitrary subdivision of $X$ into $X^1$ and $X^2$ we can compute $\varphi$ as follows,

$$\varphi[\tau, X^1, X^2] = I(X_{t-\tau}, X_t) - I(X^1_{t-\tau}, X^1_t) - I(X^2_{t-\tau}, X^2_t)$$

[2]

Thus, is $\varphi$ computed as the mutual information (see Eq. 1) between the present state of the system and its state $\tau$ time units before, with the same computation subtracted for the two subsystems $X^1$ and $X^2$.

After exhaustively computing $\varphi$ for all bipartitions, the one yielding the lowest value of $\varphi$ is identified as the minimum information bipartition (MIB). The integrated information reflects how well the system can predict its future beyond its MIB. This results in $\Phi = \varphi[\tau, X^1_*, X^2_*]$, with the MIB = $\{X^1_*, X^2_*\}$. A normalization factor is required, which is computed as the minimum of the entropies of $X^1_*$ and $X^2_*$.

This definition of $\Phi$ is understood as measuring the minimum amount of information that is lost by splitting the system into two sub-systems. For instance, a large value of $\Phi$ implies that the information generated by the system is greatly diminished by splitting it into sub-systems.

By finding the peak of the autocorrelation function of the dynamics generated by the model, we set the parameter $\tau$ = 3 time steps.

*2.7 Modular decomposition of the anatomical network*

The exhaustive evaluation of all bipartitions is intractable for systems comprising a large number of units. It would be computationally intractable to compute $\Phi$ for the 998 time series generated by our model. We therefore clustered the nodes based on their modular membership within the anatomical connectivity network. For this we applied the Louvain algorithm (using 100 iterations and keeping the highest modularity value) to the weighted adjacency matrix $W_{ij}$ to reveal six modules (Blondel et al., 2008). These modules comprised nodes tightly connected between them, but loosely connected with nodes belonging to the other modules.

After averaging the time series generated by the model (setting again 1 for active and 0 for inactive or refractory), we obtained six signals to which we could apply the procedure described in Eq. 2 for all possible ($2^6$) bipartitions, and then obtain the value of $\Phi$. This is understood to be an approximation to the real Φ of the system; however, it has been proven a useful approximation in previous work (Sasai et al., 2016; Mediano et al., 2016).

*2.8 Metastability*

We understand by metastability (ξ) the repertoire of configurations that the system explores throughout its temporal evolution. As a simple example based on a physical system, consider a ball in a landscape comprised of a series of peaks and troughs. Without sufficient kinetic energy, the system will display a simple behavior (i.e. the ball will be stuck oscillating at the bottom of a through). However, as energy is increased, the system can explore a wider range of configurations, since the ball can reach the points of metastability associated with other peaks and troughs (Tagliazucchi et al., 2016).

To extend this analogy to our system, we quantified the level of global cohesion of the average time series from the six modules (i.e. mean functional connectivity, computed using Pearson's correlation coefficient). Using non-overlapping windows of 20 time steps, we obtained the evolution of global cohesion in time. We identified the metastability with the variance of this time series. Thus, a low variance implied that the system remained homogeneously coupled over time, whereas higher values of variance indicated an exploration of more heterogeneous connectivity patterns (see Fraiman & Chialvo, 2012 and Haimovici et al., 2013 for a similar development).

## 3. Results

*3.1 Critical thresholds and network connectivity.*

We first determined the critical threshold for self-sustained activity ($T_c$) for activity generated at each of the 998 nodes of the network. In **Figure 1A** we show all network nodes (at their corresponding anatomical locations) and we highlighted in red those nodes with the highest 25% $T_c$ values, and in blue those with the lowest 25% $T_c$ values. Activity originating from the red nodes becomes self-sustained even with relatively high threshold values, whereas that originating from the blue nodes can be extinguished with relatively low values of the threshold.

We then investigated the relationship between $T_c$ and node strength. The strength of the i-th node is defined as the sum of the weights of all connections attached to it, i.e. $\sum_j W_{ij}$. It is clear from **Figure 1B** that weakly connected nodes presented low $T_c$ values. As node strength increased, so did $T_c$, up to a point at which $T_c \approx 0.45$, saturating independently of node strength.

In **Figure 1C** we show examples of the spatio-temporal propagation of activity starting from a node with $T_c \approx 0.45$. The vertical axis indexes the 998 nodes of the network (ordered by first activation time) and the temporal evolution is encoded in the horizontal axis. The color code represents the probability that a node becomes activated over a total of 1000 independent simulations. It is evident that a low propagation threshold resulted in a quickly decaying feedforward "sweep" of activity. A high propagation threshold resulted in a very high probability of observing activated nodes at any moment in time. Finally, the critical threshold $T_C$ allowed for self-sustained activity with intermediate levels of activity after the initial feedforward "sweep".

We show the average activity generated by the system for different thresholds in **Figure 1D**. To keep the system steadily generating dynamical behavior, one node was kept constantly activated throughout the simulations. Dynamics for a threshold < $T_C$ presented epochs of quiescence (dominated by a large proportion of inhibition in the form of nodes in the refractory state) alternating with bursts of activity (a pattern qualitatively similar to that of burst suppression, observed in the electroencephalogram of unconscious patients) (Hofmeijer et al., 2014). On the other hand, dynamics at $T_C$ consisted of more complex fluctuations of irregular amplitude.

*3.2 Definition of regions of interest*

Since an exhaustive investigation of all 998 nodes in the network would be unpractical, we proceeded to investigate the properties of activity propagating from a reduced set of regions of interest (ROI). These ROI were located at primary sensory (left and right visual and auditory) areas, and were chosen as the nodes initially activated in all following simulations. This choice of ROI reflects the fact that conscious access of sensory information is a process starting with the activation of neurons at primary sensory cortices. The visual and auditory ROI were located by their proximity to the calcarine sulcus and the Heschl's gyrus, respectively, identified using the automated anatomical labeling template (AAL) (Tzourio-Mazoyer et al., 2002).

The location of the ROI is shown in **Figure 2A** (V1 (L/R): left/right primary visual, A1 (L/R): left/right primary auditory). In **Figure 2B** we show the probability for sustained activity (over 1000 trials) as a function of the threshold for each of ROI. The inset displays the variability in the probability, peaking at ≈ 0.045 and coinciding with the point of highest slope in the plot of self-sustained activity probability vs. threshold. We therefore took $T_C$ = 0.045 for all four ROI.

*3.3 Mutual information and algorithmic complexity*

We then proceeded to investigate the average mutual information shared by the simulated activity as a function of the propagation threshold. Following the work of King and colleagues (King et al., 2013), we divided the analysis into short and long-range information sharing. The matrix of Euclidean distances between all pairs of nodes is shown in **Figure 3A**. The average mutual information <MI> for short-range connections was computed as the average mutual information between pairs of regions separated by less than 80 mm (approximately the mean value of the entries in the matrix of **Figure 3A**), and is shown in **Figure 3B** (left panel). The same analysis performed for long-range connections is shown in **Figure 3C** (right panel). In both cases, we observed that for almost all ROI, the average mutual information between the model activity time series peaked at $T_C$, i.e. at the point at which activity originating from the ROI became self-sustained and percolated to the whole network.

The analysis of algorithmic complexity (Lempel-Ziv complexity) as a function of threshold is shown in **Figure 3C**. A peak in the mean algorithmic complexity of the activity generated from all ROI was observed close to $T_C$, but at a slightly higher threshold value.

*3.4 Identification of nodes with highest mean mutual information*

In **Figure 4A**, we show the matrices of mutual information between all pairs of signals for three different thresholds (0.06, 0.045 and 0.02) and for all ROI. For all ROI, a non-homogeneous and structured pattern of mutual information values could be seen only at the critical threshold $T_c$ = 0.045.

For each node, we computed its mean information sharing value, and highlighted the 25% nodes with the highest information sharing values for each ROI in the right panel of **Figure 4A** (the ROI where activity originated are indicated with a colored arrow). While the patterns clearly depend on the ROI, commonalities exists, which are highlighted in **Figure 4B**. This figure shows the nodes that are at the intersection of all the top 25% nodes for each ROI; in other words, they are the nodes that consistently appeared as information sharing hubs irrespectively of the ROI where activity originated. These nodes could be found in the posterior cingulate cortex (precuneus) and in the bilateral angular gyri in the parietal cortex.

*3.5 Information integration (Φ) and metastability (ξ)*

The results of the modularity analysis on the anatomical connectivity network are shown in **Figure 5A**, with nodes color-coded based on their module membership.

We extracted the mean signal from all nodes in each module and computed Φ between them as a function of the threshold. We observed that, for all regions of interest, Φ peaked close to the point at which activity became self-sustained ($T_C$) (**Figure 5B**).

The plot of metastability (ξ) as a function of the threshold presented a similar shape for all ROI, but peaked at a slightly lower value than $T_C$.

*3.6 Serial competing stimuli*

We now turn to the question of whether competitive dynamics can qualitatively account for the results of psychophysical experiments designed to probe conscious access. We first turn to the paradigm known as backward masking, in which a mask is first presented to the participants and, a short time afterwards, an image (e.g. a face) is also presented (Breitmeyer & Ogmen, 2000). The time elapsed between the presentation of the mask and the image is known as the stimulus onset asynchrony (SOA). For low values of SOA the second image is not consciously perceived, whereas longer SOA (>100 ms) guarantee that both the mask and the image are consciously perceived. This empirical result can be interpreted according to the global workspace theory as the mask inhibiting access to the global workspace for short SOA due to the competitive nature of both incoming stimuli (Del Cul et al., 2007).

To model such effect, we serially activated the same ROI with different delays between both activations (the equivalent of the SOA in the backward masking paradigm). In this case, both activations were competitive in nature, and both propagated at $T_c$ = 0.045.

For all ROI we observed the same qualitative behavior. The probability of the second activation becoming self-sustained was low if the delay relative to the first activation was short. However, the likelihood of the second activation percolating through the network increased with the delay between serial activations (**Figure 6**). This analysis shows that activity can be globally broadcasted when presented serially at the same ROI, but only if a sufficiently long time is elapsed between the first and the second activations. It qualitatively matches psychophysical plots of conscious perception probability vs. SOA, such as those shown in Figure 2 of Del Cul et al., 2007.

*3.7 Parallel competing stimuli*

We now consider competing stimuli simultaneously presented at different ROI. To model differential stimuli strength or attention, each of them was modeled using its own propagation threshold. This simulation is useful to account for the results of paradigms such as binocular rivalry, in which two different images are independently presented to each eye, and only one of them is consciously perceived at any given moment in time (Tong et al., 1998).

**Figure 7A** presents the probability of self-sustained activity as a function of each of the two thresholds. For instance, the first panel presents the threshold for activity generated in left V1 in the horizontal axis, and for right V1 in the vertical axis; with the probability of self-sustained activity originating from right V1 color-coded in the matrix. These results confirm the intuition that if a low threshold is used for activity originating from one of the ROI, this activity will percolate and extinguish the activity originating from the other competing ROI.

The rightmost panel of **Figure 7A** shows that, depending on the relative value of the thresholds, activity originating from left V1 can become self-sustained (red area) or that originating from right V1 (blue area). However, we did not observe a region in which both activations could jointly reach a self-sustained regime, i.e. they could not both simultaneously percolate through the whole brain network. A similar result was observed for left and right A1 (**Figure 7B**). We can interpret these results as a dichotomous access to the global network, like that observed in visual or auditory rivalry (Brancucci & Tommasi, 2011) paradigms.

## 4. Discussion

In this paper we introduced a conceptual computational model simulating the propagation of activity from seed ROI to the global anatomical network. Identifying such global propagation with conscious access, we found that low threshold values resulted in quickly decaying activity, and that a node-dependent critical threshold $T_c$ existed, at which activity became self-sustained and achieved global and reverberating percolation. Under the hypothesis that large-scale propagation of activity in the brain is tuned to this critical point, maximal or close to maximal values of mutual information sharing, algorithmic complexity, information integration ($\Phi$) and metastability ($\xi$) were observed. Our modeling effort therefore establishes a link between the predictions of the information integration theory (built upon the phenomenology of consciousness) and the global workspace theory (a mechanistic explanation of conscious access). The fact that self-sustained activity propagation at $T_c$ eventually guarantees a re-activation of the chosen ROI also resonates with theories of conscious awareness based on recurrent information propagation (Lamme & Roelfsema, 2000). Furthermore, we show that the introduction of competitive dynamics in the model can qualitatively account for the results of psychophysical experiments, based on the restricted serial and parallel access to the global workspace.

They key hypothesis for establishing this integration is that of activity propagation at $T_c$. Previous work by Wallace has suggested that conscious access can be understood as the non-linear transition towards a giant connected component in the network of functional connections (Wallace, 2005). This work is an important precedent of our study, but does not incorporate dynamics and is therefore unable to link the transition with observables such as the complexity measures mentioned in the preceding paragraph. Similarly, recent work by Misic and colleagues investigated the diffusion of activity from seed ROI throughout the anatomical connectome. Their model is an important inspiration for the present work, but lacks dynamics once the diffusion process is complete and thus cannot investigate the complexity of the propagated activity during a steady state (Misic et al., 2015).

Another important antecedent of our work is the theory of neuropercolation, a mathematical theory of probabilistic cellular automata on networks aiming to reproduce the micro- and mesoscale dynamics of the brain (Kozma, 2007). The existence of phase transitions in neuropercolation models (determined by $T_c$ in our model) can explain the complexity observed at different spatial and temporal scales in the brain. A model based on neuropercolation has recently been employed to simulate loss of conscious awareness in general anesthesia; displacement from the critical point towards a point $> T_c$ leads to EEG patterns typical of unconscious patients (such as those presented in **Figure 1D**) (Zhou et al., 2015). Our work extends these efforts by directly linking conscious access with percolation of activity, and providing proof that critical propagation is a sufficient condition for the emergence of the complex dynamics characteristic of the conscious state.

It must be emphasized that the criticality hypothesis is helpful to explain the properties of large-scale information routing in the thalamo-cortical system associated with the conscious state, but lacks (at least so far) explanatory power for the neural computations performed at specific, local brain circuits. The structural networks of the brain serve as the backbone for the dynamic coordination of information processed in distributed circuits, prompting the question of how this dynamic coordination can effectively take place between the extremes of overwhelming levels of brain activity and a weak percolation of activity throughout the network (Chialvo, 2010). The hypothesis that incoming sensory stimuli propagate at $T_c$ provides a natural solution for this conundrum. However, an account of the contents of a given conscious experience are in principle beyond the scope of our hypothesis, which only concerns the conditions under which conscious access takes place in the human brain.

The mathematical framework developed by Tononi and colleagues in a series of seminal papers represents a leading effort in the quest to quantify the level of consciousness of any physical system. Such framework is often referred to as a "thermometer" for consciousness, i.e. a way to assign a numerical value ($\Phi$) to the level of consciousness. This characterization, however, does not full justice to the goal of this research programme, since different "thermometers" of consciousness have been proposed before (consider for instance the bispectral index, employed in anesthesia monitoring) (Myles et al., 2004). A more accurate description would be that of a metric based on theoretical principles that can gauge the level of consciousness in an absolute scale, akin to a Kelvin thermometer in physics (i.e. a scale measuring temperature relative to the absolute zero).

In spite of these efforts, the theory does not propose a mechanism by which a given physical system (such as the human brain) can achieve high values of Φ. It could be argued that millions of years of evolution led to the development of the cortico-thalamic complex, a system of astonishing and irreducible complexity able to produce equally complex dynamics. The criticality hypothesis offers the alternative (and simplifying) perspective that universal physical principles can lead to complex dynamics such as those observed in the human brain (Chialvo, 2010). A wealth of experimental data is concordant with this hypothesis, including scale-free avalanches observed at different spatial scales (Beggs & Plenz, 2003, Tagliazucchi et al., 2012; Shriki et al., 2013), long-range correlations in brain dynamics (Linkenkauer-Hansen et al., 2001; He, 2011), and divergence of correlation length (Fraiman & Chialvo, 2012; Haimovici et al., 2013), among others.

If proven correct, the criticality hypothesis offers a view on emergent brain complexity alternative to that of an irreducible complexity intrinsic to the biological structure of the cerebral cortex. Its predictions are of particular relevance in the context of different states of consciousness. Physical systems at criticality present a maximal susceptibility, i.e. a maximal response to external perturbations (Tagliazucchi et al., 2016). This property resonates with the observation of complex and sustained neuroelectrical responses to transcraneal magnetic stimulation during conscious wakefulness – which could be indicative of critical dynamics – as opposed to rapidly vanishing transients during unconscious states (Casali et al., 2013). The critical state also guarantees that a system can access the widest range of its functional repertoire (Chialvo, 2010; Deco & Jirsa, 2012; Tagliazucchi et al., 2016). We have shown that conscious access at $T_c$ guarantees both propagation to the whole anatomical network, but also the subsequent exploration of the possible functional configurations of the system (**Figure 5C**). Loss of signatures of consciousness have been reported in unconscious states such as general anesthesia (Scott et al., 2014; Tagliazucchi et al., 2016), epileptic seizures (Meisel et al., 2012) and deep sleep (Priesemann et al., 2013; Tagliazucchi et al., 2013). The theoretical work of Gerhard Werner suggested an early link between critical dynamics and consciousness (Werner, 2013).

A model of consciousness should have predictive power concerning the neural correlates of any conscious experience. Our computational model identified a posterior (precuneus / parietal) set of information sharing hubs that could be identified with a component of the global workspace. Dehaene and colleagues have proposed that frontal areas of the brain also contribute to the global workspace (Dehaene, 2014). However, recent experiments using no-report paradigms (Aru et al., 2012; Tsuchiya et al., 2015) and within-state studies (Siclari et al., 2014; Koch et al., 2016b) converge towards a posterior "hotspot" as a key anatomical region for conscious access. The predictions of our model are most likely influenced by the underlying structural network topology (i.e. structural hubs presenting the highest amount of information sharing) and the relative position of the hubs and the ROI in the network, suggesting a future evaluation of anatomical regions putatively involved in conscious access in terms of their connectivity and topological distance to primary sensory areas.

Consisting of a simplified view of activity propagation in the brain, our model faces several limitations. We did not include inhibition specifically in the model, even though the refractory state effectively acted as an inhibitory mechanism. We did not disambiguate attention from stimulus strength and therefore were unable to simulate conditions at the brink of conscious access (i.e. preconscious access) (Dehaene et al., 2006). We also did not provide an explanation of why activity propagation is tuned to $T_c$, even though we speculate that a self-organizing mechanism involving plasticity promotes an intermediate regime between insufficient levels of information transmission, and an "explosive" activation regime which could lead to epileptic seizures (Hellyer et al., 2016). The simplicity of our model, on the other hand, allows the identification of the critical activity propagation as a possible key component behind the emergent complexity of brain dynamics during conscious access. More neurobiologically oriented models have been constructed before, including realistic equations for neuronal behavior and different cell types (Dehaene et al., 2003). These models provided a quantitatively accurate picture of the ignition of the global workspace, but required the tuning of a larger number of parameters, and their activity is difficult to put in parametric correspondence with observables derived from the information integration theory.

## 5. Conclusion

In conclusion, we proposed that two influential theories could be compatible when considered as addressing different aspects of consciousness (i.e. conscious access vs. the phenomenology of consciousness). While the motivation behind the development of both theories might be different, the criticality hypothesis offers a picture in which experimental predictions from both of them can successfully co-exist. It is therefore important to focus future experimental and theoretical efforts on testing such hypothesis, especially whether incoming sensory information propagates at the edge of becoming self-sustained.

**Conflict of interest**

The research was conducted in the absence of any commercial or financial relationships that could be construed as a potential conflict of interest.

**Acknowledgments**

The author acknowledges insightful discussions with Pablo Balenzuela, Dante Chialvo, Ariel Haimovici, Helmut Laufs, Maximiliano Zeller, Pablo Gonzalez, and thanks Olaf Sporns for sharing the DSI connectome data. E.T. was supported by a postdoctoral fellowship of the AXA Research Fund.

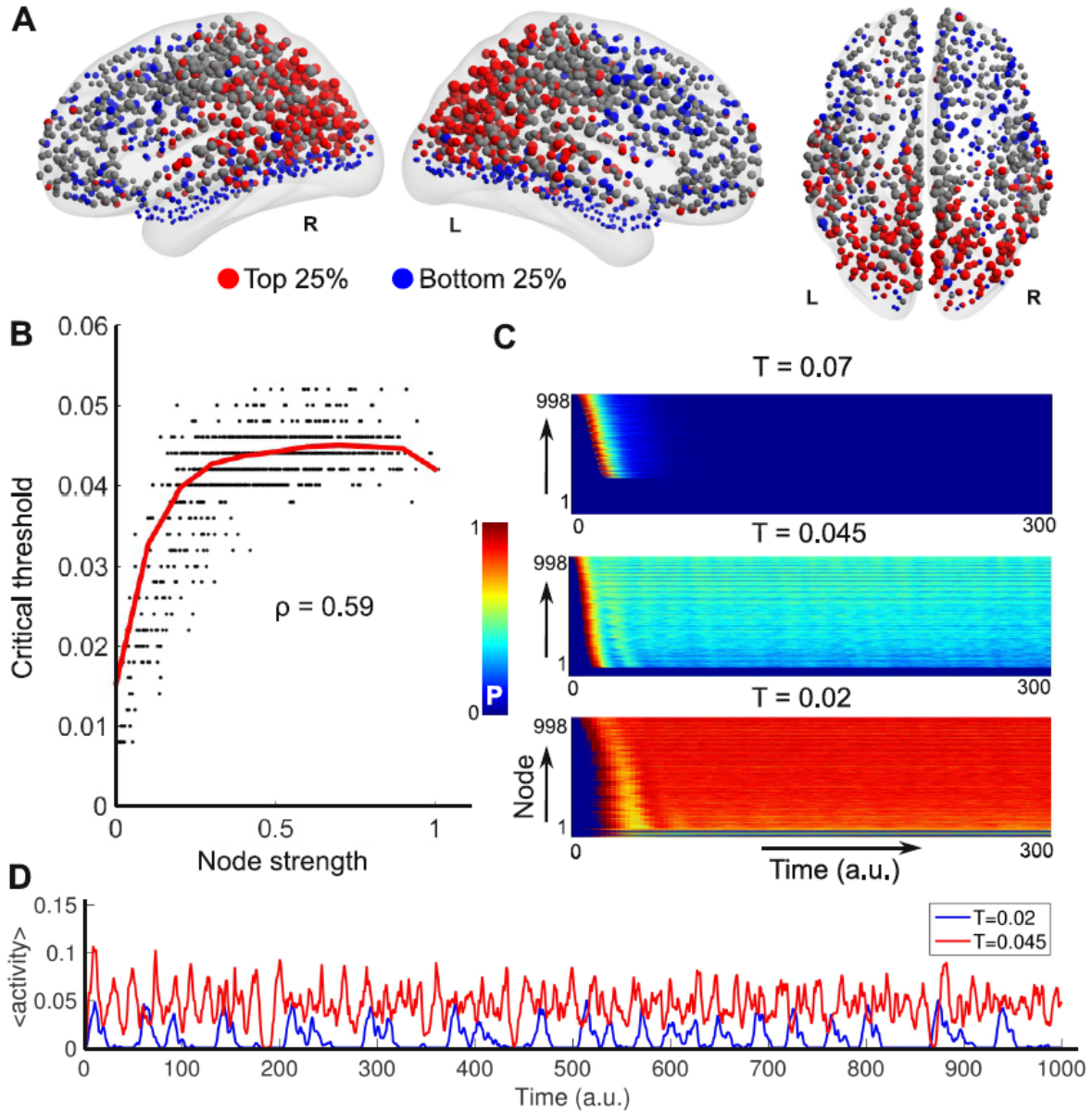

**Figure 1**: *Critical thresholds for sustained activity, examples of model dynamics and relationship to anatomical connectivity*. **A.** Anatomical rendering of 998 nodes, nodes with the 25% highest thresholds are shown in red, those with the bottom 25% in blue. **B.** Critical threshold vs. node strength for all 998 nodes (the red line represents the mean). **C.** Probabilities of observing node activations (over 1000 independent simulations) as a function of time for three different thresholds. **D.** Dynamics of the mean activity of the system with one node permanently activated driving the system, and for two different threshold values.

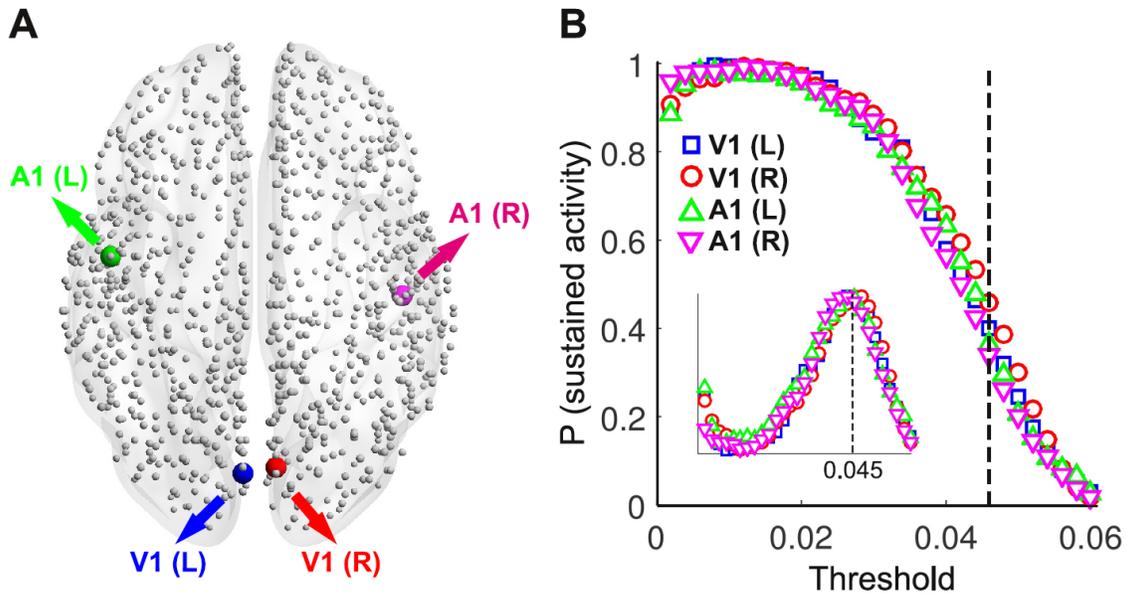

**Figure 2**: Definition of r*egions of interest (ROI)*. **A.** Four nodes corresponding to the ROI investigated throughout our work (left and right primary auditory and visual cortices). **B.** Probability of eliciting self-sustained activity as a function of the threshold for activity propagation. Different symbols represent results for activity starting from each ROI. The inset shows the variability as a function of the threshold, and the vertical dashed line indicates the critical point ($T_C$).

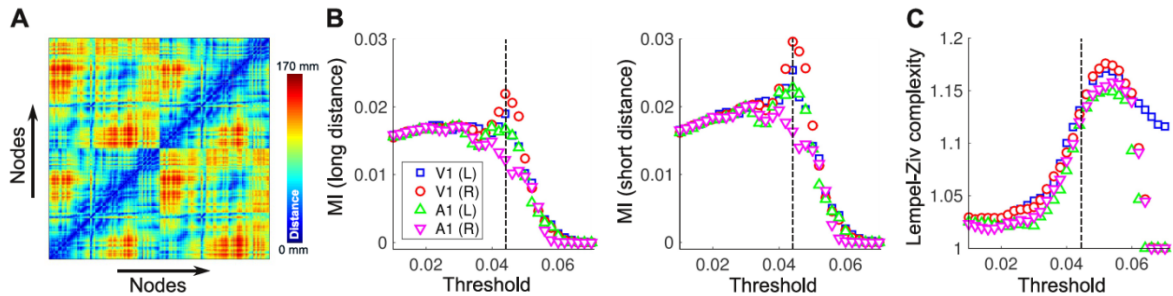

**Figure 3**: *Information sharing and algorithmic complexity are maximal close to $T_C$*. **A.** Matrix of Euclidean distance between all pairs of nodes in the anatomical connectivity network. **B.** Average mutual information between the time series of the nodes, considering only long-distance connections (left) and short-distance connections (right). Different symbols indicate activity elicited from each ROI. **C.** Average Lempel-Ziv complexity of all time series for activity elicited at each ROI as a function of the threshold. In all cases, the vertical dashed line indicates $T_C$.

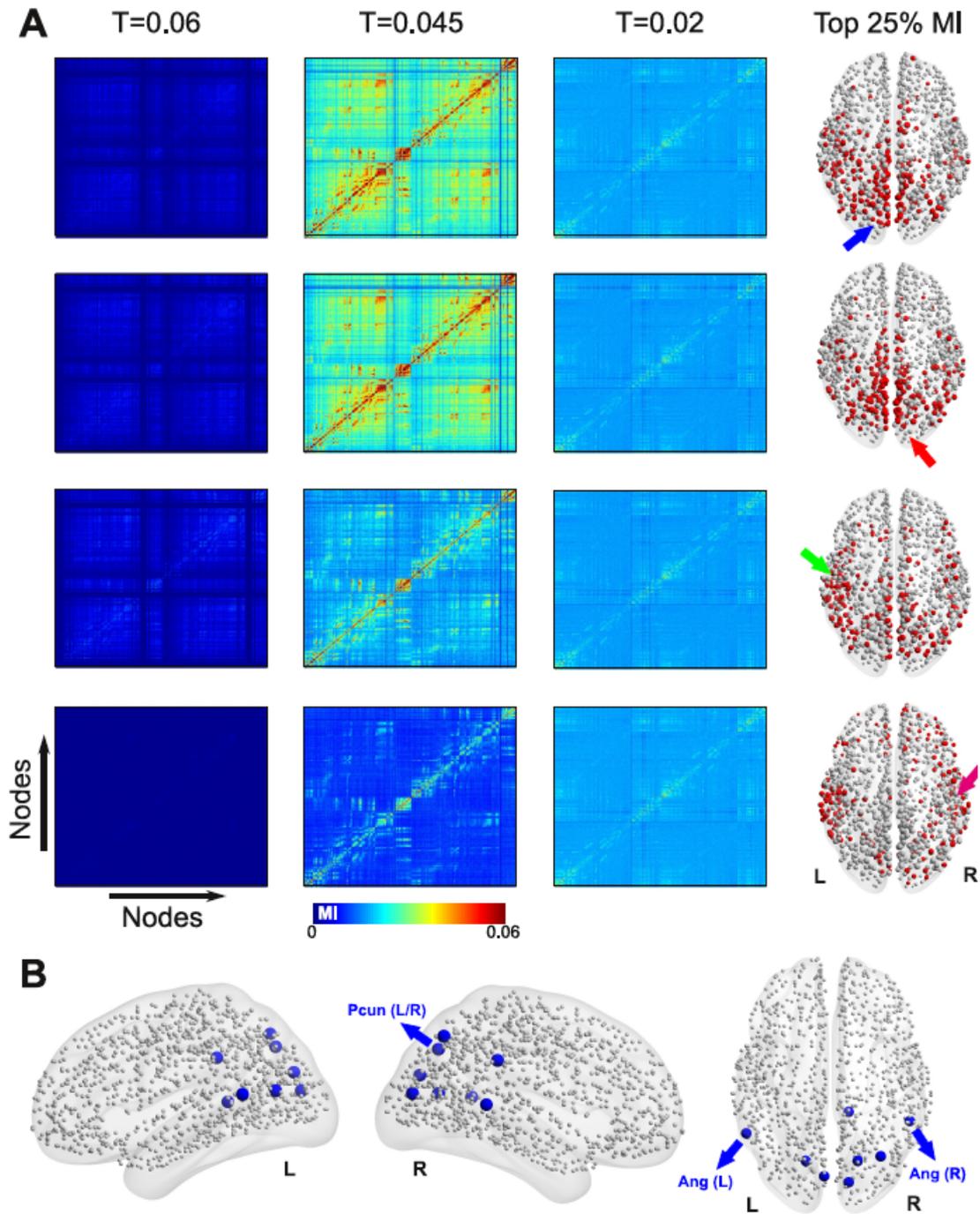

**Figure 4**: *Nodes associated with the highest information integration are located at posterior cingulate and parietal regions.* **A.** Mean mutual information matrices for three thresholds (middle column corresponds to $T_C$). Each row corresponds to activity elicited at a different ROI (left and right V1, and left and right A1, respectively). In the rightmost column the nodes associated with the top 25% highest average information sharing are marked in red. **B.** Intersection between all of the top 25% nodes with highest average information sharing (rightmost column of panel A). Pcun: precuneus, Ang: angular gyrus.

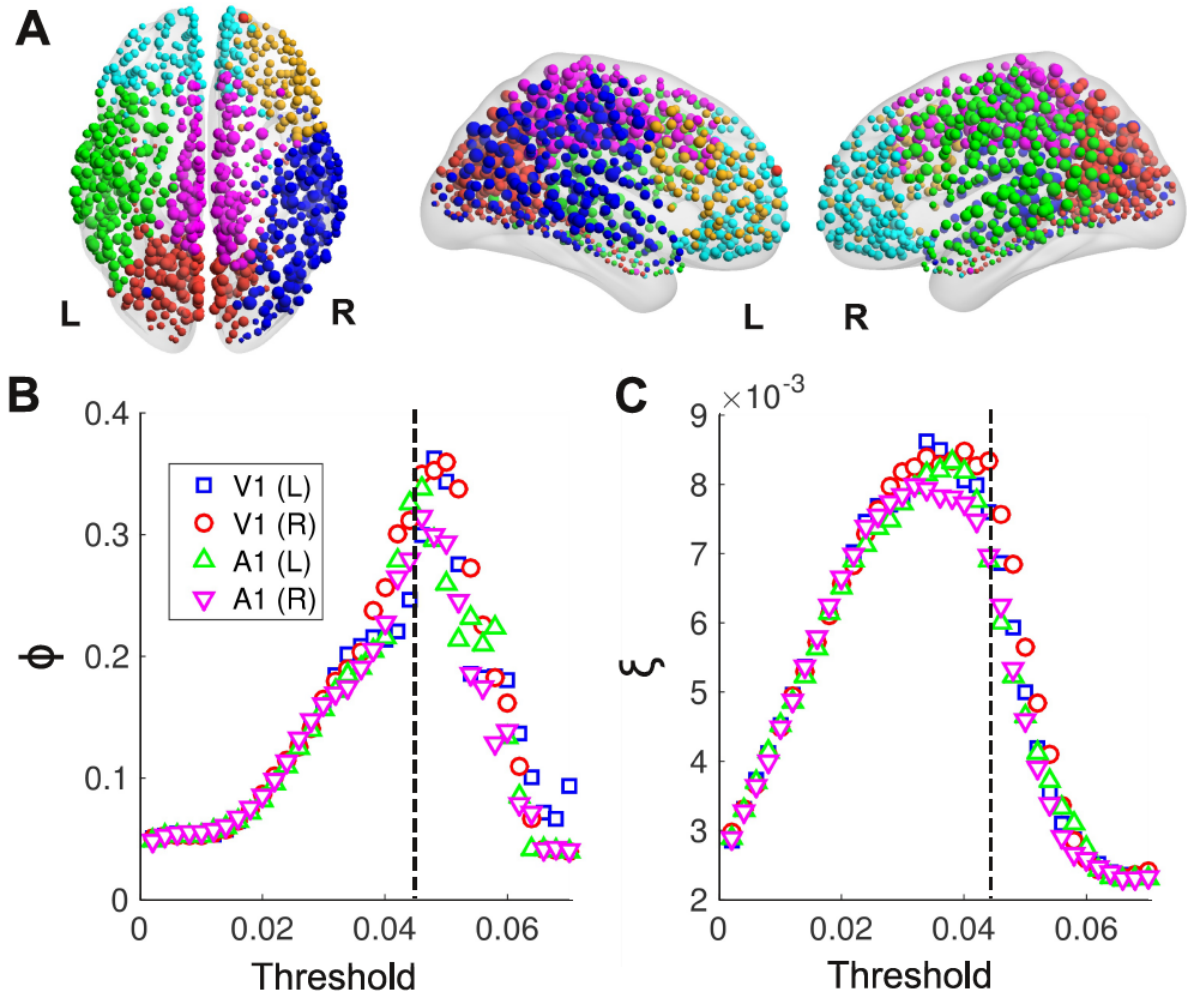

**Figure 5**: *Maximal information integration (Φ) and metastability (ξ) peak close to $T_C$.* **A.** Modules of the anatomical connectivity network determined using the Louvain algorithm. Nodes belonging to each of the six modules are represented in a different color. **B.** Information integration (Φ) as a function of the threshold; different symbols represent activity elicited at each ROI. Φ was computed between the six time series of average activity obtained averaging the model activity within each of the modules presented in Panel A. **C.** Metastability (ξ) as a function of the threshold of the model, different symbols represent activity elicited at each ROI. ξ was computed as the temporal variance of the mean correlation matrices obtained from the six time series comprising the model dynamics averaged at each node presented in Panel A.
23

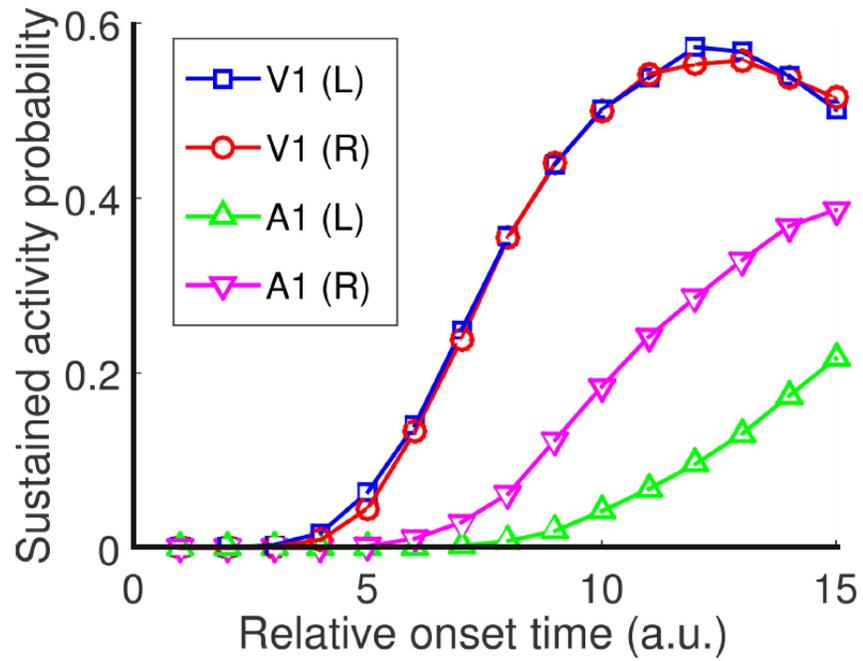

**Figure 6**: *Competing serial activations*. Probability of observing sustained activity resulting from a second activation, after first activating each ROI, plotted as a function of the relative onset time of both serial activations.

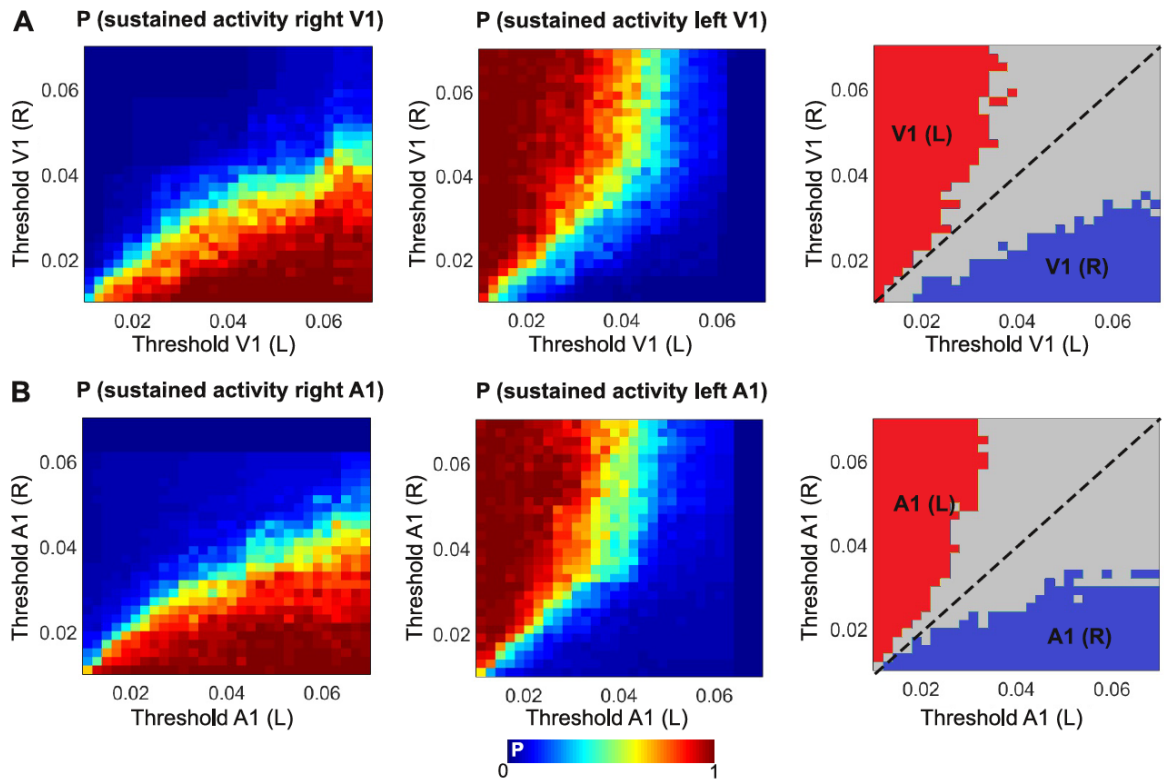

**Figure 7**: *Competing parallel activations*. **A.** Probability of observing sustained activity for activity generating in right V1 (right column) and left V1 (middle column) as a function of the propagation threshold of the activities simultaneously elicited at left and right V1. The rightmost panel shows the threshold combinations allowing self-sustained activity for activations originating from left and right V1 (notice the lack of overlap). **B.** Same results as in Panel A, but using the ROI in the left and right A1 instead.